\documentclass[11pt]{article}
\usepackage{amsmath}
\usepackage{graphicx}
\usepackage{amsfonts}
\usepackage{amssymb}
\usepackage{stmaryrd}
\usepackage{amscd}
\newcommand{\bc}{\begin{center}}
\newcommand{\ec}{\end{center}}
\newcommand{\be}{\begin{equation}}
\newcommand{\ee}{\end{equation}}
\newcommand{\bea}{\begin{eqnarray}}
\newcommand{\eea}{\end{eqnarray}}
\newcommand{\ba}{\begin{array}}
\newcommand{\ea}{\end{array}}
\newcommand{\edc}{\end{document}}

\newcommand{\ol}{\overline}

\def\f{\varphi}

\def\O{\Omega}

\def\s{\sigma}

\def\C{{\cal C}}

\begin{document}
\thispagestyle{empty}
\begin{center}

{\bf DESCRIPTION OF WEAK PERIODIC GROUND STATES OF ISING MODEL WITH COMPETING INTERACTIONS ON CAYLEY TREE}\\
\vspace{0.4cm}
M.M.Rahmatullaev \footnote{mrahmatullaev@rambler.ru}\\
{\it Namangan State University, 316, Uychi str., Namangan,
 Uzbekistan}\\

\vspace{0.5cm}

{\bf Abstract}

\end{center}

Recently by Rozikov an Ising  model with competing interactions
and  spin values $\pm 1$, on a Cayley tree of order $k\geq 1$ has
been considered and the ground states of the model are described.
In this paper we describe some weak periodic ground states of the
model.

{\bf Keywords:} Cayley tree, configuration, Ising model, ground
state.

\section{Introduction}

\large This paper gives a class of weak periodic ground states for
an Ising  model with competing interactions and  spin values $\pm
1$, on a Cayley tree of order $k\geq 1$.  One of the key problems
related to the spin models is the description of the set of Gibbs
measures. This problem has a good connection with the problem of
the description the set of ground states. Because the phase
diagram of Gibbs measures (see [7], [12] for details) is close to
the phase diagram of the ground states for sufficiently small
temperatures. Usually, more simple and interesting ground states
are periodic ones. But for some set of parameters such  a ground
state does not exist. In such a case it would be necessary to find
some a weak periodic ground states.

The Ising model, with two values of spin $\pm 1$  was considered
in [10],[13] and became actively researched in the 1990's and
afterwards (see for example  [1]-[6], [8], [9], [11]).

 \textbf{The Cayley tree}. The Cayley tree $\Gamma^k$ (See [2]) of
order $ k\geq 1 $ is an infinite tree, i.e., a graph without
cycles, from each vertex of which exactly $ k+1 $ edges issue. Let
$\Gamma^k=(V, L, i)$ , where $V$ is the set of vertices of $
\Gamma^k$, $L$ is the set of edges of $ \Gamma^k$ and $i$ is the
incidence function associating each edge $l\in L$ with its
endpoints $x,y\in V$. If $i(l)=\{x,y\}$, then $x$ and $y$ are
called {\it nearest neighboring vertices}, and we write $l=<x,y>$.

 The distance $d(x,y), x,y\in V$ on the Cayley tree is defined by the formula
$$
d(x,y)=\min\{ d | \exists x=x_0,x_1,...,x_{d-1},x_d=y\in V \ \
\mbox{such that}  \ \
 <x_0,x_1>,...,<x_{d-1},x_d> \}.$$

For the fixed $x^0\in V$ we set $ W_n=\{x\in V\ \ |\ \
d(x,x^0)=n\},$
$$ V_n=\{x\in V\ \ | \ \  d(x,x^0)\leq n\}, \ \
L_n=\{l=<x,y>\in L \ \ |\ \  x,y\in V_n\}. \eqno (1) $$
 Denote $|x|=d(x,x^0)$, $x\in V$.

A collection of the pairs $<x,x_1>,...,<x_{d-1},y>$ is called a
{\sl path} from $x$ to  $y$ and we write $\pi(x,y)$ .
 We write $x<y$ if
the path from $x^0$ to $y$ goes through $x$.

It is known (see [6]) that there exists a one-to-one
correspondence between the set  $V$ of vertices  of the Cayley
tree of order $k\geq 1$ and the group $G_{k}$ of the free products
of $k+1$ cyclic  groups $\{e, a_i\}$, $i=1,...,k+1$ of the second
order (i.e. $a^2_i=e$, $a^{-1}_i=a_i$) with generators $a_1,
a_2,..., a_{k+1}$.

Denote $S(x)$ the set of "direct successors" of $x\in G_k$. Let
$S_1(x)$ be denotes the set of all nearest neighboring vertices of
$x\in G_k,$ i.e. $S_1(x)=\{y\in G_k: <x,y>\}$ and $ x_{\downarrow}
=S_1(x)\setminus S(x)$.

\textbf{ The model.} Here we shall give main definitions and facts
about the model which we are going to study (see [1] for details).
Consider models where the spin takes values in the set
$\Phi=\{-1,1\}$. For $A\subseteq V$ a spin {\it configuration}
$\s_A$ on $A$ is defined as a function
 $x\in A\to\s_A(x)\in\Phi$; the set of all configurations coincides with
$\Omega_A=\Phi^{A}$. Denote $\O=\O_V$ and $\s=\s_V.$ Also put
$-\s_A=\{-\s_A(x), x\in A\}.$  Define a {\it periodic
configuration} as a configuration $\s\in \O$ which is invariant
under a subgroup of shifts $G^*_k\subset G_k$ of finite index.
More precisely, a configuration $\s \in \Omega $ is called $G^*_k$
-periodic if $\s (yx)=\s (x)$ for any $x\in G_k$ and $y\in G^*_k$.

  For a given periodic configuration  the index of the subgroup is
called the {\it period of the configuration}. A configuration
 that is invariant with respect to all
shifts is called {\it translational-invariant}. Let
$G_k/G_k^*=\{H_1,...,H_r\}$  factor group, where $G_k^*$ is a
normal subgroup of index $r\geq 1$. Configuration $\sigma(x), x\in
V$ is called  $G_k^*$ {\it weak
 periodic}, if $\sigma(x)=\sigma_{ij}$ for  $x\in H_i, x_\downarrow\in H_j, \forall x\in
 G_k$.

The Hamiltonian  of the Ising model with competing interactions
has the form
$$
H(\s)=J_1 \sum\limits_{<x,y>}\s(x)\s(y)+ J_2 \sum\limits_{x,y\in
V: \ \ d(x,y)=2}{\s(x)\s(y)}  \eqno (2)
$$
where $J_1, J_2\in R$ are coupling constants and $\s\in \Omega$.

For a pair of configurations $\s$ and $\f$ that coincide almost
everywhere, i.e. everywhere except for a finite number of
positions, we consider a relative Hamiltonian $H(\s,\f)$, the
difference between the energies of the configurations $\s, \f$ of
the form
$$
H(\s,\f)=J_1\sum_{<x,y>}(\sigma(x)\sigma(y)-\f(x)\f(y))+
 J_2\sum_{x,y\in V: d(x,y)=2}(\sigma(x)\sigma(y)-\f(x)\f(y)), \eqno (3)
$$
where $J=(J_1,J_2)\in R^2$ is an arbitrary fixed parameter.

Let $M$ be the set of unit balls  with vertices in $V$. We call
the restriction of a configuration $\s$ to the ball $b\in M$ a
{\it bounded configuration} $\s_b .$

Define the energy of a ball $b$ for configuration $\s$ by
$$
U(\s_b)\equiv U(\s_b, J)=\frac{1}{2}J_1 \sum\limits_{<x,y>,\ \
x,y\in b} \s(x)\s(y)+ J_2 \sum\limits_{x,y\in b:\ \
d(x,y)=2}{\s(x)\s(y)},  \eqno (4)
$$
where $J=(J_1, J_2)\in R^2.$

We shall say that two bounded configurations $\s_b$ and $\s'_{b'}$
belong to the same class if $U(\s_b)=U(\s'_{b'})$ and we write
 $\s'_{b'}\sim \s_b. $

For any set $A$ we denote by $|A|$ the number of elements in $A$.

\vskip 0.5 truecm

{\bf Lemma 1.} [1] {\it 1) For any configuration $\s_b$ we have}
$$U(\s_b)\in \{U_0, U_1, ..., U_{k+1}\},$$
{\it where
$$U_i=\bigg(\frac{k+1}{2}-i\bigg)J_1+\bigg(\frac{k(k+1)}{2}+2i(i-k-1)\bigg)J_2,
\ \ i=0,1,...,k+1. \eqno(5)$$

2) Let $\C_i=\O_i\cup \O^-_i,\ \  i=0,...,k+1,$ where
$$\O_i=\big\{\s_b: \s_b(c_b)=+1,\ \  |\{x\in b\setminus \{c_b\}:
 \s_b(x)=-1\}|=i\big\},$$
$$\O^-_i=\big\{-\s_b=\{-\s_b(x), x\in b\}: \s_b\in \O_i \big\},$$
and $c_b$ is the center of the ball \ $b$. Then for $\s_b\in \C_i$
we have $U(\s_b)=U_i.$}

3) {\it The class $\C_i$ contains $\frac{2(k+1)!}{i!(k-i+1)!}$
configurations.}

\vskip 0.5 truecm

{\bf Definition 1.} A  configuration $\f$ is called a {\it ground
state} for the relative Hamiltonian $H$ if
$$ U(\f_b)=\min\{U_0, U_1,..., U_{k+1}\},\ \  \mbox{ for any}
\ \  b\in M. \eqno (6)$$

\vskip 0.2 truecm

We set
$$U_i(J)=U(\s_b,J), \ \ \mbox{if} \ \ \s_b\in \C_i,\ \  i=0,1,...,k+1.$$

The quantity $U_i(J)$ is a linear function of the parameter $J\in
R^2. $ For every fixed $m=0,1,...,k+1$ we denote

$$A_m=\{J\in R^2: U_m(J)=\min\{U_0(J), U_1(J),..., U_{k+1}(J)\}\}. \eqno (7)$$

It is easy to check that
$$A_0=\{J\in R^2: J_1\leq 0;\ \  J_1+2kJ_2\leq 0\}; $$
$$A_m=\{J\in R^2: J_2\geq 0;\ \  2(2m-k-2)J_2\leq J_1
\leq 2(2m-k)J_2\}, \ \ m=1,2,...,k; $$
$$A_{k+1}=\{J\in R^2: J_1\geq 0;\ \  J_1-2kJ_2\geq 0\} $$
and $R^2=\cup_{i=0}^{k+1}A_i .$

For any $A_i, A_j, i\ne j$ we have
$$A_i\cap A_j=\left\{\begin{array}{lll}
\{J: J_1=2(2i-k)J_2, \ \ J_2\geq 0\}& \textrm{if \ \
$ j=i+1, \ \ i=0,1,...,k$}\\
(0,0)& \textrm{if \ \ $1<|i-j|<k+1$}\\
\{J: J_1=0, J_2\leq 0\}& \textrm{if \ \ $|i-j|=k+1$}\\
\end{array}\right. \eqno(8)$$

Denote
$$B=A_0\cap A_{k+1}, \ \ B_i=A_i\cap A_{i+1},\ \  i=0,...,k.$$
$${\tilde A}_0=A_0\setminus (B\cup B_0), \ \
{\tilde A}_{k+1}=A_{k+1}\setminus (B\cup B_k),$$
$$ {\tilde A}_i=A_i\setminus (B_{i-1}\cup B_i),\ \ i=1,...,k.$$

Fix $J\in R^2$ and denote
$$N_J(\s_b)=|\{j: \s_b\in \C_j\}|.$$

Let $GS(H)$ be the set of all ground states of the relative
Hamiltonian $H$ (see (4)).

For any $\s=\{\s(x), \ \ x\in V\}\in \O$ denote ${\ol
\s}=-\s=\{-\s(x), \ \ x\in V\}.$

In work [1] the set of periodic ground states for the model (1) is
described i.e. the following is proved.

 \vskip 0.2 truecm

 {\bf Theorem 1.} {\it (i) If $J=(0,0)$ then
$GS(H)=\O$.

(ii) If $J\in {\tilde A}_i, \ \ i=0,...,k+1$ then
$$GS(H)=\{\s^{(i)}, {\ol \s}^{(i)}\}.$$

(iii) If $J\in B_i\setminus \{(0,0)\}, \ \ i=0,...,k$ then
$$GS(H)=\{\s^{(i)}, {\ol \s}^{(i)},\s^{(i+1)}, {\ol \s}^{(i+1)}\}\cup
S_i,$$ where $S_i$ contains at least a countable subset of  non
periodic ground states.

(iv) If $J\in B\setminus \{(0,0)\},$ then
$$GS(H)=\{\s^{(0)}, {\ol \s}^{(0)},\s^{(k+1)}, {\ol \s}^{(k+1)}\}.$$

Here $\s^{(i)}, \ \ {\ol \s}^{(i)},\ \ i=0,...,k+1$ are periodic
ground states such that on any $b\in M$ the bounded configurations
$\s^{(i)}_b, {\ol \s}^{(i)}_b \in \C_i,$ i.e. $\s^{(0)}, {\ol
\s}^{(0)}$ are translational - invariant and  $\s^{(i)}, {\ol
\s}^{(i)}, \ \ i=1,...,k+1$ are periodic with period 2.}

 \textbf{Remark 1.} We note, that weak periodic (non periodic) ground states belong to
the set $S_i$ i.e. for parameters $J_1=2(2i-k)J_2, J_2\neq 0$ .

In this paper we explicitly describe the weak periodic (with
respect to normal subgroups of index 2 and 4) ground states.

\section{Weak periodic ground states.}

\textbf{Case:index 2.}

Let $A\subset \{1,2,...,k+1\}$, $H_A=\{x\in G_k: \sum_{j\in
A}w_j(x)-$even$\},$ where $w_j(x)$-is the number of letters $a_j$
in the word $x.$

It is obvious, that $H_A$ is a normal subgroup of index two [6].
Let $G_k/H_A=\{H_A,G_k\setminus H_A\}$ be the quotient group.

We set $H_0=H_A, H_1=G_k\setminus H_A$.

The  $H_A$ - weak periodic configurations have the form:

(1)  $
\varphi_1(x)=\pm \left\{%
\begin{array}{ll}
    +1, & {x_{\downarrow} \in H_0} \ x \in H_0 \\
    +1, & {x_{\downarrow} \in H_0} \ x \in H_1 \\
    +1, & {x_{\downarrow} \in H_1} \ x \in H_0 \\
    +1, & {x_{\downarrow} \in H_1} \ x \in H_1, \\
\end{array}%
\right. $ \qquad (2)  $
\varphi_2(x)=\pm \left\{%
\begin{array}{ll}
    -1, & {x_{\downarrow} \in H_0} \ x \in H_0 \\
    +1, & {x_{\downarrow} \in H_0} \ x \in H_1 \\
    +1, & {x_{\downarrow} \in H_1} \ x \in H_0 \\
    +1, & {x_{\downarrow} \in H_1} \ x \in H_1, \\
\end{array}%
\right. $
\\
\\

(3)  $
\varphi_3(x)=\pm \left\{%
\begin{array}{ll}
    +1, & {x_{\downarrow} \in H_0} \ x \in H_0 \\
    -1, & {x_{\downarrow} \in H_0} \ x \in H_1 \\
    +1, & {x_{\downarrow} \in H_1} \ x \in H_0 \\
    +1, & {x_{\downarrow} \in H_1} \ x \in H_1, \\
\end{array}%
\right. $ \qquad (4)  $
\varphi_4(x)=\pm \left\{%
\begin{array}{ll}
    +1, & {x_{\downarrow} \in H_0} \ x \in H_0 \\
    +1, & {x_{\downarrow} \in H_0} \ x \in H_1 \\
    -1, & {x_{\downarrow} \in H_1} \ x \in H_0 \\
    +1, & {x_{\downarrow} \in H_1} \ x \in H_1, \\
\end{array}%
\right. $
\\
\\

(5)  $
\varphi_5(x)=\pm \left\{%
\begin{array}{ll}
    +1, & {x_{\downarrow} \in H_0} \ x \in H_0 \\
    +1, & {x_{\downarrow} \in H_0} \ x \in H_1 \\
    +1, & {x_{\downarrow} \in H_1} \ x \in H_0 \\
    -1, & {x_{\downarrow} \in H_1} \ x \in H_1, \\
\end{array}%
\right. $ \qquad (6)  $
\varphi_6(x)=\pm \left\{%
\begin{array}{ll}
    -1, & {x_{\downarrow} \in H_0} \ x \in H_0 \\
    -1, & {x_{\downarrow} \in H_0} \ x \in H_1 \\
    +1, & {x_{\downarrow} \in H_1} \ x \in H_0 \\
    +1, & {x_{\downarrow} \in H_1} \ x \in H_1, \\
\end{array}%
\right. $
\\
\\

(7)  $
\varphi_7(x)=\pm \left\{%
\begin{array}{ll}
    -1, & {x_{\downarrow} \in H_0} \ x \in H_0 \\
    +1, & {x_{\downarrow} \in H_0} \ x \in H_1 \\
    -1, & {x_{\downarrow} \in H_1} \ x \in H_0 \\
    +1, & {x_{\downarrow} \in H_1} \ x \in H_1, \\
\end{array}%
\right. $ \qquad (8)  $
\varphi_8(x)=\pm \left\{%
\begin{array}{ll}
    +1, & {x_{\downarrow} \in H_0} \ x \in H_0 \\
    -1, & {x_{\downarrow} \in H_0} \ x \in H_1 \\
    -1, & {x_{\downarrow} \in H_1} \ x \in H_0 \\
    +1, & {x_{\downarrow} \in H_1} \ x \in H_1. \\
\end{array}%
\right. $

Hence, we must choose weak periodic ground states among these 16
configurations. The following theorem gives the result.

\textbf{ Theorem 2.} {\it Let $|A|=i, i\in \{1,2,...,k+1\}$

1) If $|A|\neq \frac{k+1}{2}$ then each $H_A$-weak periodic ground
state is a $H_A$- periodic or translational-invariant i.e. belongs
to the set $\{\pm \varphi_1(x),\pm \varphi_7(x)\}$.

2) If $|A|=\frac{k+1}{2}$  then there are at least two two  $H_A$
- weak periodic (non-periodic) ground states which are of the form
$\pm \varphi_8(x).$}

\textbf{Proof.} By (8) one can see that a configuration $\phi$ is
a ground state if and only if there is $j\in \{0,...,k\}$ such
that $\phi_b\in \C_j\cup \C_{j+1}$ for any $b\in M$. Thus we must
check this property for above mentioned configurations.

1) Let

$
\varphi_1(x)= \left\{%
\begin{array}{ll}
    +1, & {x_{\downarrow} \in H_0} \ x \in H_0 \\
    +1, & {x_{\downarrow} \in H_0} \ x \in H_1 \\
    +1, & {x_{\downarrow} \in H_1} \ x \in H_0 \\
    +1, & {x_{\downarrow} \in H_1} \ x \in H_1. \\
\end{array}%
\right. $

It is obvious, that $H_A$ weak periodic ground states are
translational-invariant.

 2)Let
$
\varphi_2(x)= \left\{%
\begin{array}{ll}
    -1, & {x_{\downarrow} \in H_0} \ x \in H_0 \\
    +1, & {x_{\downarrow} \in H_0} \ x \in H_1 \\
    +1, & {x_{\downarrow} \in H_1} \ x \in H_0 \\
    +1, & {x_{\downarrow} \in H_1} \ x \in H_1. \\
\end{array}%
\right. $

$\forall b\in M$ we have $|\{x\in S_1(c_b):x\in H_0\}|=i$,
$|\{x\in
S_1(c_b):x\in H_1\}|=k+1-i.$ \\
Denote $A_{-}  =\{x\in S_1(x):\varphi_b(x)=-1\},$
$A_{+}=\{x\in S_{1}(x):\varphi_b(x)=+1\},$ and $\varphi_{i,b}=(\varphi_i)_b, i=1,2,...,8.$\\

Assume $c_b\in H_0.$ The possible cases are: \\
 a) $c_{b \downarrow} \in H_0$ and $\varphi_{2,b}(c_{b
\downarrow})=+1$, then $\varphi_{2,b}(c_{b})=-1$ è $|A_{-}|=i-1,
|A_+|=k+2-i, \varphi_{2,b} \in C_{k+2-i}.$\\
 b) $c_{b \downarrow}
\in H_0$ and $\varphi_{2,b}(c_{b \downarrow})=-1$, then
$\varphi_{2,b}(c_{b})=-1$ è $|A_{-}|=i, |A_+|=k+1-i, \varphi_{2,b}
\in
C_{k+1-i}.$ \\
c) $c_{b \downarrow} \in H_1$ and $\varphi_{2,b}(c_{b
\downarrow})=+1$, then $\varphi_{2,b}(c_{b})=+1$ è $|A_{-}|=i,
|A_+|=k+1-i, \varphi_{2,b} \in C_{i}.$

If $c_b\in H_1 $ then we have \\
d) if $c_{b \downarrow} \in H_0$ and $\varphi_{2,b}(c_{b
\downarrow})=-1$, then $\varphi_{2,b}(c_{b})=+1$ è $|A_{-}|=1,
|A_+|=k, \varphi_{2,b} \in C_{1},$ \\ e) If $c_{b \downarrow} \in
H_0$ è $\varphi_{2,b}(c_{b \downarrow})=+1$, then
$\varphi_{2,b}(c_{b})=+1$ è $|A_{-}|=0, |A_+|=k+1, \varphi_{2,b}
\in C_{0}.$

By (8) we get $C_{k+2-i}\cap C_{k+1-i}\cap C_i=\emptyset$ if
$i\neq \frac{k+1}{2},\frac{k+2}{2}.$ Now consider the case $i=
\frac{k+1}{2},$ from  d) and e) we get $i=0, k=-1,$ that is
impossible. If $i= \frac{k+2}{2},$ then $i=1, k=0,$ which also
impossible. Thus, $\varphi_2(x)$ is not a ground state.

3) Let

$
\varphi_3(x)= \left\{%
\begin{array}{ll}
    +1, & {x_{\downarrow} \in H_0} \ x \in H_0 \\
    -1, & {x_{\downarrow} \in H_0} \ x \in H_1 \\
    +1, & {x_{\downarrow} \in H_1} \ x \in H_0 \\
    +1, & {x_{\downarrow} \in H_1} \ x \in H_1. \\
\end{array}%
\right. $

Let $c_b\in H_0.$ Consider several cases: \\
a) $c_{b \downarrow} \in H_0$ and $\varphi_{3,b}(c_{b
\downarrow})=+1$, then $\varphi_{3,b}(c_{b})=+1$ è $|A_{-}|=k+1-i,
|A_+|=i, \varphi_{3,b} \in C_{k+1-i},$\\
b) $c_{b \downarrow} \in H_1$ and $\varphi_{3,b}(c_{b
\downarrow})=+1$, then $\varphi_{3,b}(c_{b})=+1$ è $|A_{-}|=k-i,
|A_+|=i+1, \varphi_{3,b} \in C_{k-i},$ \\
c) $c_{b \downarrow} \in H_1$ and $\varphi_{3,b}(c_{b
\downarrow})=-1$, then $\varphi_{3,b}(c_{b})=+1$ and
$|A_{-}|=k+1-i, |A_+|=i, \varphi_{3,b} \in C_{k+1-i}.$

Let $c_b\in H_1$. We have: \\
d) $c_{b \downarrow} \in H_0$ and $\varphi_{3,b}(c_{b
\downarrow})=+1$, then $\varphi_{3,b}(c_{b})=-1$ and $|A_{-}|=0,
|A_+|=k+1, \varphi_{3,b} \in C_{k+1},$\\  e) $c_{b \downarrow} \in
H_1$ and $\varphi_{3,b}(c_{b \downarrow})=-1$, then
$\varphi_{3,b}(c_{b})=+1$ and $|A_{-}|=1, |A_+|=k, \varphi_{3,b}
\in C_{1},$ \\ f) $c_{b \downarrow} \in H_1$ è $\varphi_{3,b}(c_{b
\downarrow})=+1$, then $\varphi_{3,b}(c_{b})=+1$ è $|A_{-}|=0,
|A_+|=k+1, \varphi_{3,b} \in C_{0}.$

By (8) $C_{0}\cap C_{1}\cap C_{k+1}=\emptyset$ if $k\neq 0.$
 Thus, $\varphi_3(x)$ is not a ground state.

4) For $\varphi_j(x), \ \ j=4,5,6$ one similarly can prove that
they are not ground states.

5) Consider now

$
\varphi_7(x)= \left\{%
\begin{array}{ll}
    -1, & {x_{\downarrow} \in H_0} \ x \in H_0 \\
    +1, & {x_{\downarrow} \in H_0} \ x \in H_1 \\
    -1, & {x_{\downarrow} \in H_1} \ x \in H_0 \\
    +1, & {x_{\downarrow} \in H_1} \ x \in H_1 \\
\end{array}%
\right.
= \left\{%
\begin{array}{ll}
    -1, & {x \in H_0} \\
    +1, & {x \in H_1.} \\
\end{array}%
\right. $

Consequently  $\varphi_7(x)$ is a periodic ground state which is
not interesting for us.

6) Consider

$
\varphi_8(x)= \left\{%
\begin{array}{ll}
    +1, & {x_{\downarrow} \in H_0} \ x \in H_0 \\
    -1, & {x_{\downarrow} \in H_0} \ x \in H_1 \\
    -1, & {x_{\downarrow} \in H_1} \ x \in H_0 \\
    +1, & {x_{\downarrow} \in H_1} \ x \in H_1. \\
\end{array}%
\right. $

Let $c_b\in H_0.$ The possible cases are \\
a) $c_{b \downarrow} \in H_0$ and $\varphi_{8,b}(c_{b
\downarrow})=+1$, then $\varphi_{8,b}(c_{b})=+1$ è $|A_{-}|=k+1-i,
|A_+|=i, \varphi_{8,b} \in C_{k+1-i},$\\  b)   $c_{b \downarrow}
\in H_0$ and $\varphi_{8,b}(c_{b \downarrow})=-1$, then
$\varphi_{8,b}(c_{b})=+1$ and $|A_{-}|=k+2-i, |A_+|=i-1,
\varphi_{8,b} \in C_{k+2-i}$, \\ c) $c_{b \downarrow} \in H_1$ and
$\varphi_{8,b}(c_{b\downarrow})=-1$, then
$\varphi_{8,b}(c_{b})=-1$ and $|A_{-}|=k+1-i, |A_+|=i,
\varphi_{8,b} \in C_{i},$ \\ d) $c_{b \downarrow} \in H_1$ and
$\varphi_{8,b}(c_{b\downarrow})=+1$, then
$\varphi_{8,b}(c_{b})=-1$ and $|A_{-}|=k-i, |A_+|=i+1,
\varphi_{8,b} \in C_{i+1}.$

For $c_b\in H_1$ we have \\
e) $c_{b \downarrow} \in H_0$ and $\varphi_{8,b}(c_{b
\downarrow})=+1$, then $\varphi_{8,b}(c_{b})=-1$ and $|A_{-}|=k-i,
|A_+|=i+1, \varphi_{8,b} \in C_{i+1},$\\  f)   $c_{b \downarrow}
\in H_0$ and $\varphi_{8,b}(c_{b \downarrow})=-1$, then
$\varphi_{8,b}(c_{b})=-1$ and $|A_{-}|=k+1-i, |A_+|=i,
\varphi_{8,b} \in C_{i}$, \\ g) $c_{b \downarrow} \in H_1$ and
$\varphi_{8,b}(c_{b\downarrow})=-1$, then
$\varphi_{8,b}(c_{b})=+1$ and $|A_{-}|=k+2-i, |A_+|=i-1,
\varphi_{8,b} \in C_{k+2-i},$ \\ h) $c_{b \downarrow} \in H_1$ and
$\varphi_{8,b}(c_{b\downarrow})=+1$, then
$\varphi_{8,b}(c_{b})=+1$ and $|A_{-}|=k+1-i, |A_+|=i,
\varphi_{8,b} \in C_{k+1-i}.$

We obtain $C_{i}\cap C_{i+1}\cap C_{k+1-i}\cap
C_{k+2-i}=\emptyset$ if $\left\{%
\begin{array}{ll}
    i\neq k+1-i \\
    i+1\neq k+2-i \\
\end{array}%
\right. $ or  $\left\{%
\begin{array}{ll}
    i\neq k+2-i \\
    i+1\neq k+1-i. \\
\end{array}%
\right. $ Only following system has solution
$$\left\{%
\begin{array}{ll}
    i=k+1-i \\
    i+1=k+2-i \\
\end{array}%
\right. \Rightarrow i=\frac{k+1}{2}.$$

Thus  the configuration  $\varphi_8$ is a ground state iff
$i=\frac{k+1}{2}$. The theorem is proved.

\textbf{Corollary.} {\it If $k$ is a even number then each weak
periodic ground state is a periodic one.}

\textbf{Case: index 4.}

Take $A\subset \{1,2,...,k+1\}$. $H_A=\{x\in G_k: \sum_{j\in
A}w_j(x)-$even$\},$ $G^{(2)}_k=\{x\in G_k:|x|$-even $\}$ where
$w_j(x)$ is the number of $a_j$ in word $x.$ $G^{(4)}_k=H_A\cap
G^{(2)}_k-$ normal subgroup of index 4 [6].

$G_k/G^{(4)}_k=\{H_0,H_1,H_2,H_3\},$ where \\
$ H_0=\{x\in G_k: \sum_{j\in A}w_j(x)$-even$, |x|$-even $\},$\\
$ H_1=\{x\in G_k: \sum_{j\in A}w_j(x)$-odd$, |x|$-even$\},$\\
$ H_2=\{x\in G_k: \sum_{j\in A}w_j(x)$-even$, |x|$-odd $\},$\\
$ H_3=\{x\in G_k: \sum_{j\in A}w_j(x)$-odd$, |x|$-odd$\}.$\\

$G^{(4)}_k-$ weak periodic configuration has the form\\
$$
\varphi(x)= \left\{%
\begin{array}{ll}
    a_{13}, & {x_{\downarrow} \in H_1} \ x \in H_3 \\
    a_{31}, & {x_{\downarrow} \in H_3} \ x \in H_1 \\
    a_{03}, & {x_{\downarrow} \in H_0} \ x \in H_3 \\
    a_{30}, & {x_{\downarrow} \in H_3} \ x \in H_0 \\
    a_{21}, & {x_{\downarrow} \in H_2} \ x \in H_1 \\
    a_{12}, & {x_{\downarrow} \in H_1} \ x \in H_2 \\
    a_{02}, & {x_{\downarrow} \in H_0} \ x \in H_2 \\
    a_{20}, & {x_{\downarrow} \in H_2} \ x \in H_0 \\
\end{array}%
\right.\eqno(9) $$ where $a_{pq}=\pm 1, p,q\in\{0,1,2,3\}.$

Thus we have to determine which of them are ground states. The
result is

\textbf{Theorem 3.} {\it Let $|A|=i, i\in \{1,2,...,k+1\}$.\\
i) If $i\neq \frac{k+1}{2}$ then each $G^{(4)}_k$-weak periodic
ground state is a periodic.\\
ii) If $i=\frac{k+1}{2}$, then there are periodic and four
$G^{(4)}_k$-weak periodic ground states:\\
$
\pm \varphi'(x)= \left\{%
\begin{array}{ll}
    +1, & {x_{\downarrow} \in H_1} \ x \in H_3 \\
    +1, & {x_{\downarrow} \in H_3} \ x \in H_1 \\
    -1, & {x_{\downarrow} \in H_0} \ x \in H_3 \\
    -1, & {x_{\downarrow} \in H_3} \ x \in H_0 \\
    -1, & {x_{\downarrow} \in H_2} \ x \in H_1 \\
    -1, & {x_{\downarrow} \in H_1} \ x \in H_2 \\
    +1, & {x_{\downarrow} \in H_0} \ x \in H_2 \\
    +1, & {x_{\downarrow} \in H_2} \ x \in H_0, \\
\end{array}%
\right.$ and $
\pm \varphi''(x)= \left\{%
\begin{array}{ll}
    -1, & {x_{\downarrow} \in H_1} \ x \in H_3 \\
    +1, & {x_{\downarrow} \in H_3} \ x \in H_1 \\
    +1, & {x_{\downarrow} \in H_0} \ x \in H_3 \\
    -1, & {x_{\downarrow} \in H_3} \ x \in H_0 \\
    -1, & {x_{\downarrow} \in H_2} \ x \in H_1 \\
    +1, & {x_{\downarrow} \in H_1} \ x \in H_2 \\
    -1, & {x_{\downarrow} \in H_0} \ x \in H_2 \\
    +1, & {x_{\downarrow} \in H_2} \ x \in H_0. \\
\end{array}%
\right.$}

\textbf{Remark 2.}

a) Using Theorems 1-3 we can give the phase diagram of the ground
states for any $k$.

b) By Remark 1 and our theorems we get $J_1=2J_2,J_2>0.$
Consequently, for ordinary Ising model (i.e. $J_2=0$) there is no
weak periodic ground state. Since for $J_1=2J_2=0$ the Hamiltonian
equals to zero.

c) For cases of index other than 2 and 4 the description of weak
periodic ground states becames a technically difficult problem.

d) We note that any normal subgroup of index two has a form $H_A$
for a suitable $A\subset \{1,2,...,k+1\}.$ But there are many
normal subgroups of index four which do not coincide with
$G_k^{(4)}$, for example, $H_A\cap H_B$ for $A,B\varsubsetneqq
\{1,2,...,k+1\}$ with $A\neq B$ is a normal subgroup of index four
but it does not coincide with $G_k^{(4)}.$

\textbf{Proof of Theorem 3.} Consider several cases of
configurations (9).

1) Let $a_{pq}=+1 (a_{pq}=-1), \forall p,q\in\{0,1,2,3\}.$
Obviously, $G^{(4)}_k$-weak periodic ground states are translation
invariant.

2) $\forall b\in M$  we have \\
$|\{x\in S_1(c_b): c_b\in H_0, x\in H_3\}|=i$, $|\{x\in S_1(c_b):
c_b\in H_0, x\in H_2\}|=k+1-i$,\\
$|\{x\in S_1(c_b): c_b\in H_1, x\in H_2\}|=i$, $|\{x\in S_1(c_b):
c_b\in H_1, x\in H_3\}|=k+1-i$,\\
$|\{x\in S_1(c_b): c_b\in H_2, x\in H_1\}|=i$, $|\{x\in S_1(c_b):
c_b\in H_2, x\in H_0\}|=k+1-i$,\\
$|\{x\in S_1(c_b): c_b\in H_3, x\in H_0\}|=i$, $|\{x\in S_1(c_b):
c_b\in H_3, x\in H_1\}|=k+1-i$.\\

Denote $A_{-}  =\{x\in S_1(x):\varphi_b(x)=-1\},$
$A_{+}=\{x\in S_{1}(x):\varphi_b(x)=+1\}.$\\

Put $a_{13}=-1,$ for others  $a_{pq}=+1.$\\
Assume $c_b\in H_0,$ then the following cases are possible:\\
a) $c_{b\downarrow}\in H_3$ and
$\varphi_b(c_{b\downarrow})=a_{13}=-1,$ then
$\varphi_b(c_{b})=a_{30}=+1, |A_{-}|=1, |A_{+}|=k,$ consequently $\varphi_b \in C_1,$\\
b) $c_{b\downarrow}\in H_3$ and
$\varphi_b(c_{b\downarrow})=a_{03}=+1,$ then
$\varphi_b(c_{b})=a_{30}=+1, |A_{-}|=0, |A_{+}|=k+1,$ thus
$\varphi_b \in C_0,$\\
If $c_b\in H_3,$ then we have\\
c) $c_{b\downarrow}\in H_1$ and
$\varphi_b(c_{b\downarrow})=a_{31}=+1,$ then
$\varphi_b(c_{b})=a_{13}=-1, |A_{-}|=0, |A_{+}|=k+1,$ consequently
$\varphi_b \in C_{k+1}.$

By (8) we one can see that a configuration $\phi$ is a ground
state if and only if there is $j\in \{0,...,k\}$ such that
$\phi_b\in \C_j\cup \C_{j+1}$ for any $b\in M$. Thus it is enough
to check this property for above mentioned configurations.
 We have $C_0\cap C_1 \cap C_{k+1}\neq \emptyset $ for any
$k\geq 1.$ Thus $\varphi$ is not a ground state. Similarly one can
prove that if $p_0,q_0, a_{p_0q_0}=-1$ and others $a_{pq}=+1$,
then the configuration is not a ground state.

3) Let $a_{13}=a_{31}=-1$ and others $a_{pq}=+1$.\\
If $c_b\in H_0$, we have\\
a) $c_{b\downarrow}\in H_3$ and
$\varphi_b(c_{b\downarrow})=a_{13}=-1,$ then
$\varphi_b(c_{b})=a_{30}=+1, |A_{-}|=1, |A_{+}|=k,$ thus
$\varphi_b \in C_1,$\\
b) $c_{b\downarrow}\in H_3$ and
$\varphi_b(c_{b\downarrow})=a_{03}=+1,$ then
$\varphi_b(c_{b})=a_{30}=+1, |A_{-}|=0, |A_{+}|=k+1,$ sequences
$\varphi_b \in C_0,$\\
If $c_b\in H_1$ then\\
c) $c_{b\downarrow}\in H_3$ and
$\varphi_b(c_{b\downarrow})=a_{03}=+1,$ then
$\varphi_b(c_{b})=a_{31}=-1, |A_{-}|=k-i, |A_{+}|=i+1,$ thus
$\varphi_b \in C_{i+1},$\\
d) $c_{b\downarrow}\in H_3$ and
$\varphi_b(c_{b\downarrow})=a_{13}=-1,$ then
$\varphi_b(c_{b})=a_{31}=-1, |A_{-}|=k-i+1, |A_{+}|=i,$
consequently $\varphi_b \in C_{i},$\\
e) $c_{b\downarrow}\in H_2$ and
$\varphi_b(c_{b\downarrow})=a_{12}=+1,$ then
$\varphi_b(c_{b})=a_{21}=+1, |A_{-}|=k-i+1, |A_{+}|=i,$ thus
$\varphi_b \in C_{k-i+1},$\\
By (8) $C_0\cap C_1\cap C_i\cap C_{i+1} \cap C_{k+1}\neq \emptyset
$ for all  $k\geq 1.$ Thus $\varphi$ is not a ground state. All
other cases can be checked similarly.

Now we shall prove (ii) for configurations $\varphi'$.

Let $a_{13}=a_{31}=a_{02}=a_{20}=-1$ others $a_{pq}=+1.$\\
If $c_b\in H_0$, we consider the cases:\\
a) $c_{b\downarrow}\in H_3$ and
$\varphi'_b(c_{b\downarrow})=a_{13}=-1,$ then
$\varphi'_b(c_{b})=a_{30}=+1, |A_{-}|=k+2-i, |A_{+}|=i-1,$ thus
$\varphi'_b \in C_{k+2-i},$\\
b) $c_{b\downarrow}\in H_3$ and
$\varphi'_b(c_{b\downarrow})=a_{03}=+1,$ thus
$\varphi'_b(c_{b})=a_{30}=+1, |A_{-}|=k+1-i, |A_{+}|=i,$ thus
$\varphi'_b \in C_{k+1-i},$\\
c) $c_{b\downarrow}\in H_2$ and
$\varphi'_b(c_{b\downarrow})=a_{02}=-1,$ then
$\varphi'_b(c_{b})=a_{20}=-1, |A_{-}|=k+1-i, |A_{+}|=i,$
consequently
$\varphi'_b \in C_{i}.$\\
d) $c_{b\downarrow}\in H_2$ and
$\varphi'_b(c_{b\downarrow})=a_{12}=+1,$ then
$\varphi'_b(c_{b})=a_{20}=-1, |A_{-}|=k-i, |A_{+}|=i+1,$
consequently
$\varphi'_b \in C_{i+1}.$\\
Assume $c_b\in H_1$, then\\
a1) $c_{b\downarrow}\in H_3$ and
$\varphi'_b(c_{b\downarrow})=a_{03}=+1,$ then
$\varphi'_b(c_{b})=a_{31}=-1, |A_{-}|=k-i, |A_{+}|=i+1,$
consequently
$\varphi'_b \in C_{i+1},$\\
b1) $c_{b\downarrow}\in H_3$ and
$\varphi'_b(c_{b\downarrow})=a_{13}=-1,$ then
$\varphi'_b(c_{b})=a_{31}=-1, |A_{-}|=k+1-i, |A_{+}|=i,$
consequently
$\varphi'_b \in C_{i},$\\
c1) $c_{b\downarrow}\in H_2$ and
$\varphi'_b(c_{b\downarrow})=a_{12}=+1,$ then
$\varphi'_b(c_{b})=a_{21}=+1, |A_{-}|=k+1-i, |A_{+}|=i,$ thus
$\varphi'_b \in C_{k+1-i}.$\\
d1) $c_{b\downarrow}\in H_2$ and
$\varphi'_b(c_{b\downarrow})=a_{02}=-1,$ then
$\varphi'_b(c_{b})=a_{21}=+1, |A_{-}|=k+2-i, |A_{+}|=i-1,$ thus
$\varphi'_b \in C_{k+2-i}.$\\
If $c_b\in H_2$, we consider:\\
a2) $c_{b\downarrow}\in H_0$ and
$\varphi'_b(c_{b\downarrow})=a_{20}=-1,$ then
$\varphi'_b(c_{b})=a_{02}=-1, |A_{-}|=k+1-i, |A_{+}|=i,$ thus
$\varphi'_b \in C_{i},$\\
b2) $c_{b\downarrow}\in H_0$ and
$\varphi'_b(c_{b\downarrow})=a_{30}=+1,$ then
$\varphi'_b(c_{b})=a_{02}=-1, |A_{-}|=k-i, |A_{+}|=i+1,$ hence
$\varphi'_b \in C_{i+1},$\\
c2) $c_{b\downarrow}\in H_1$ and
$\varphi'_b(c_{b\downarrow})=a_{21}=+1,$ then
$\varphi'_b(c_{b})=a_{12}=+1, |A_{-}|=k+1-i, |A_{+}|=i,$ thus
$\varphi'_b \in C_{k+1-i}.$\\
d2) $c_{b\downarrow}\in H_1$ and
$\varphi'_b(c_{b\downarrow})=a_{31}=-1,$ then
$\varphi'_b(c_{b})=a_{12}=+1, |A_{-}|=k+2-i, |A_{+}|=i-1,$
consequently
$\varphi'_b \in C_{k+2-i}.$\\
Suppose $c_b\in H_3$, then\\
a3) $c_{b\downarrow}\in H_0$ and
$\varphi'_b(c_{b\downarrow})=a_{20}=-1,$ then
$\varphi'_b(c_{b})=a_{03}=+1, |A_{-}|=k+2-i, |A_{+}|=i-1,$
consequently
$\varphi'_b \in C_{k+2-i},$\\
b3) $c_{b\downarrow}\in H_0$ è
$\varphi'_b(c_{b\downarrow})=a_{30}=+1,$ then
$\varphi'_b(c_{b})=a_{03}=+1, |A_{-}|=k+1-i, |A_{+}|=i,$ thus
$\varphi'_b \in C_{k+1-i},$\\
c3) $c_{b\downarrow}\in H_1$ and
$\varphi'_b(c_{b\downarrow})=a_{21}=+1,$ then
$\varphi'_b(c_{b})=a_{13}=-1, |A_{-}|=k-i, |A_{+}|=i+1,$ thus
$\varphi'_b \in C_{i+1}.$\\
d3) $c_{b\downarrow}\in H_1$ and
$\varphi'_b(c_{b\downarrow})=a_{31}=-1,$ then
$\varphi'_b(c_{b})=a_{13}=-1, |A_{-}|=k-i+1, |A_{+}|=i,$ hence
$\varphi'_b \in C_{i}.$\\
We have $C_{i}\cap C_{i+1}\cap C_{k+1-i}\cap
C_{k+2-i}=\emptyset$ if $\left\{%
\begin{array}{ll}
    i\neq k+1-i \\
    i+1\neq k+2-i \\
\end{array}%
\right. $ or  $\left\{%
\begin{array}{ll}
    i\neq k+2-i \\
    i+1\neq k+1-i. \\
\end{array}%
\right. $ Thus it is easy to see that the configuration is a
ground state iff $i=\frac{k+1}{2}$. By analogue one can prove that
$\varphi''(x)$ is a ground state iff $i=\frac{k+1}{2}$. The
theorem is proved. \vskip 0.3 truecm

\newpage

{\bf References}

1.U.A.Rozikov, A Constructive Description of Ground States and
Gibbs Measures for Ising Model With Two-Step Interactions on
Cayley Tree,{\it Jour. Statist. Phys.} {\bf 122} : 217-235 (2006).

2.  R.J. Baxter, {\it Exactly  Solved Models in Statistical
Mechanics}, (Academic Press, London/New York, 1982).

3. P.M. Bleher, J. Ruiz, V.A. Zagrebnov. On the purity of the
limiting Gibbs state for the Ising model on the Bethe lattice,
{\it Jour. Statist. Phys.} {\bf 79} : 473-482 (1995).

4. P.M. Bleher and  N.N. Ganikhodjaev, On pure phases of the Ising
model on the Bethe lattice, {\it Theor. Probab. Appl.} {\bf
35}:216-227 (1990) .

5. P.M. Bleher, J. Ruiz, R.H.Schonmann, S.Shlosman and V.A.
Zagrebnov, Rigidity of the critical phases on a Cayley tree, {\it
Moscow Math. Journ}. {\bf 3}: 345-362 (2001).

6.  N.N. Ganikhodjaev and U.A. Rozikov, A description of periodic
extremal Gibbs measures of some lattice models on the Cayley tree,
{\it Theor. Math. Phys.} {\bf 111}: 480-486 (1997).

7. R.A. Minlos, {\it Introduction to mathematical statistical
physics} (University lecture series,  v.19,  2000)

8. F.M. Mukhamedov , U.A. Rozikov, On Gibbs measures of models
with competing ternary and binary interactions and corresponding
von Neumann algebras.  {\it Jour. of Stat.Phys.} {\bf 114}:
825-848 (2004).

9.  Kh.A. Nazarov, U.A. Rozikov, Periodic Gibbs measures for the
Ising model with competing interactions, {\it Theor. Math. Phys.}
{\bf 135}: 881-888 (2003)

10. C. Preston, {\it Gibbs states on countable sets} (Cambridge
University Press, London 1974).

11.  U.A. Rozikov, Partition structures of the group
representation of the Cayley tree into cosets by finite-index
normal subgroups and their applications to the description of
periodic Gibbs distributions,  {\it Theor. Math. Phys.} {\bf 112}:
929-933 (1997).

12.  Ya.G. Sinai, {\it Theory of phase transitions: Rigorous
Results} (Pergamon, Oxford, 1982).

13. S. Zachary, Countable state space Markov random fields and
Markov chains on trees, {\it Ann. Prob.} {\bf 11}: 894-903 (1983).

\end{document}